\documentclass[11pt]{article}
\usepackage{graphicx}
\usepackage{amsmath}
\usepackage{amsfonts}
\usepackage{amssymb}
\usepackage{soul}
\usepackage{dsfont}
\usepackage{hyperref}
\usepackage{amstext}
\usepackage[caption=false]{subfig}
\usepackage{epstopdf} 
\usepackage{mathtools} 
\usepackage{mnsymbol} 
\usepackage{balance}
\hypersetup{colorlinks=true, citecolor=blue, urlcolor=blue, linkcolor=blue}
\usepackage{color,soul}
\usepackage{osid}
\usepackage[square,numbers]{natbib}

\usepackage{physics}
\usepackage{braket}

\newcommand{\RN}[1]{\textup{\uppercase\expandafter{\romannumeral#1}}}

\newcommand{\kket}[1]{\left|#1\right\rrangle}

\newcommand{\bbraket}[1]{\left\llangle #1 \right\rrangle}

\date{\today}

\begin{document}

\title{Exceptional points and exponential sensitivity for periodically driven Lindblad equations }

\author{Jonas Larson \\{\footnotesize\it Department of Physics, Stockholm University, AlbaNova University Center, SE-106 91 Stockholm, Sweden \& jonas.larson@fysik.su.se}\\[2ex]
        Sofia Qvarfort
                     \\{\footnotesize\it Nordita, KTH Royal Institute of Technology and Stockholm University, Hannes Alfv\'{e}ns v\"{a}g 12, SE-106 91 Stockholm, Sweden and  Department of Physics, Stockholm University, AlbaNova University Center, SE-106 91 Stockholm, Sweden \& sofia.qvarfort@fysik.su.se} }

\maketitle
\begin{abstract}
In this contribution to the memorial issue of G\"oran Lindblad, we investigate the periodically driven Lindblad equation for a two-level system. We analyze the system using both adiabatic diagonalization and numerical simulations of the time-evolution, as well as Floquet theory. Adiabatic diagonalization reveals the presence of exceptional points in the system, which depend on the system parameters. We show how the presence of these exceptional points affects the system evolution, leading to a rapid dephasing at these points and a staircase-like loss of coherence. This phenomenon can be experimentally observed by measuring, for example, the population inversion. We also observe that the presence of exceptional points seems to be related to which underlying Lie algebra the system supports.
In the Floquet analysis, we map the time-dependent Liouvillian to a non-Hermitian Floquet Hamiltonian and analyze its spectrum. For weak decay rates, we find a Wannier-Stark ladder spectrum accompanied by corresponding Stark-localized eigenstates. For larger decay rates, the ladders begin to dissolve, and new, less localized states emerge. Additionally, their eigenvalues are exponentially sensitive to perturbations, similar to the skin effect found in certain non-Hermitian Hamiltonians.\\

\noindent This paper is an invited contribution to the Lindblad memorial volume in Open Systems and Information Dynamics.
\end{abstract}

\section{Introduction} 

The Lindblad equation (LE)~\cite{lindblad1976generators}\footnote{We are
aware of the fact that the equation is often referred to as
the Gorini–Kossakowski–Sudarshan–Lindblad equation (or GKSL
for short), from the others discussing the same
type of master equation. For a contribution to a special
issue in memory of G\"oran Lindblad we take the freedom to simply call it the Lindblad equation. During a
local reception at the Royal Institute of Technogy honoring G\"oran, one of us (JL) took the opportunity to ask
G\"oran about the “GKS” part of the equation. Humble
as he was, G\"oran replied that people knew about the
general form before him, but a solid mathematical proof
was missing.} has long played a crucial role in the quantum optics community~\cite{carmichael1999statistical,scully1999quantum,breuer2002theory,gardiner2004quantum}. It is commonly used to describe the evolution of a system that unavoidably couples to the environment through the exchange of quanta. In many cases, the LE accurately captures the dynamics of the system. When the parameters of the LE are time-independent, the system's evolution is Markovian, which is often a good approximation for weak system-bath coupling and typical photon frequencies in quantum ~\cite{dann2018time}. Although non-Markovian effects can be investigated using master equations in Lindblad-form with time-dependent coefficients~\cite{tu2008non, jin2010non, lei2012quantum, zhang2012general}, our focus in this work is rather on the spectral properties of the \textit{Liouvillian}.

The LE is a linear equation that can always be vectorized to resemble Schr\"odinger's equation, where the time evolution generated by a non-Hermitian (NH) "Hamiltonian" is called the Liouvillian $\hat{\mathcal{L}}$. With the time evolution generated by a NH matrix, unitarity is broken, and moreover it may result in the appearance of exceptional points (EPs)~\cite{minganti2019quantum} as well as the skin effect~\cite{zhang2022review,okuma2020topological}.  EPs are degeneracies in the complex spectrum and correspond to the coalescence of eigenvectors. In experiments, EPs have been primarily studied in classical optical systems that mimic quantum systems with an NH Hamiltonian~\cite{miri2019exceptional,parto2020non} or NH quantum systems constrained to "post-selection"\cite{naghiloo2019quantum}. The skin effect is typically identified by two properties: all eigenvectors exponentially localize to one boundary of the system for a local finite Hamiltonian~\cite{okuma2020topological}, and the spectrum becomes exponentially sensitive to non-local perturbations~\cite{budich2020non,edvardsson2022sensitivity}. On the experimental side, the skin effect has been demonstrated in different classical systems, such as nonlinear optics~\cite{weidemann2020topological}, electrical circuits~\cite{helbig2020generalized,zou2021observation}, and ultracold atomic gases~\cite{liang2022dynamic}. All these works belong to what has become known as NH quantum mechanics (QM), which extends quantum mechanics to NH Hamiltonians~\cite{ashida2020non}. 

NH QM is a natural result of a LE analysis where the terms corresponding to "quantum jumps" are omitted. However, this seemingly simple approximation has far-reaching consequences. The time-evolution no longer preserves the norm, and the map is not CPTP (complete positivity and trace-preserving). Furthermore, it violates the \textit{no-signaling theorem}~\cite{lee2014local}, causality, LOCC~\cite{chen2014increase}, and more generally, the \textit{fluctuation-dissipation} or \textit{quantum regression theorems}~\cite{carmichael1999statistical}. Additionally, the physical interpretations of NH QM are not well-understood, which partly hinges on mathematical grounds~\cite{edvardsson2022biorthogonal}.
These issues are not present in the LE formalism, yet as mentioned before, the LE can be represented as a NH Schr\"odinger equation. Thus, EPs~\cite{mathisen2018liouvillian,minganti2019quantum,khandelwal2021signatures,minganti2019quantum} and the skin effect can still appear in the Liouvillian~\cite{song2019non,longhi2020unraveling,haga2021liouvillian,yang2022liouvillian}. However, working with the LE comes at a price - you must work in an enlarged state space, and vectors in this space may not represent physical states. As we explain below, although most eigenvectors of the Liouvillian are not physical states, they still form a complete basis for any physical state, which indeed is essential for the LE to be a dynamical CPTP map.
 
In this study, we investigate the impact of exceptional points (EP's) on a periodically dephasing qubit described by a time-periodic Lindbladian equation (LE). We show that in the absence of EP's, the qubit undergoes exponential relaxation, while in their presence, relaxation occurs in a step-like fashion. We attribute this behavior to the breakdown of adiabaticity near EP's. To gain insight into the underlying dynamics, we employ a Lie algebra decoupling method developed in the 1960s for closed system dynamics~\cite{wei1963lie} (see~\cite{qvarfort2022solving} for a tutorial). Our results suggest that exceptional points require the full algebra to generate the associated dynamics. Additionally, we explore the Floquet Hamiltonian (Liouvillian) spectrum and find a complex \textit{Wannier-Stark ladder} with eigenstates localized within the lattice. Surprisingly, a transition occurs where the ladder structure breaks down, and new types of less localized states emerge. These new eigenvalues display extreme parameter sensitivity, similar to systems that support the skin effect.

This paper is structured as follows. In Sec.~\ref{sec:tdLE}, we introduce the time-dependent Lindblad equation and discuss two commonly used parametrizations of it; regular vectorization and te Bloch representation. We also mention some general properties of the resulting Liouvillians, and discuss how the two are connected. In Sec.~\ref{sec:two:level:systems}, we analyse the periodic qubit system with the Bloch representation and using Floquet theory. The work is concluded  Sec.~\ref{sec:conclusion}. 

\section{The Lindblad equation with time-dependent coefficients, and ways of vectorisation}  \label{sec:tdLE}
The single-channel time-dependent LE is given by 
\begin{equation} \label{eq:Lindblad:general}
\dot{\hat{\rho}} = - i [\hat H, \hat \rho] + \gamma(t) \left(  2\hat L  \, \hat \rho \, \hat L^\dagger - \{ \hat L^\dagger \hat L , \hat \rho \} \right), 
\end{equation}
where $\hat\rho$ is the density matrix of the system, $\hat H$ is the system Hamiltonian, and $\hat L$ is the Lindblad operator for the single decay channel. Throughout, we will consider a time-independent Hamiltonian, while the rate $\gamma(t)=\gamma_0^{(n)}\left(1+\cos(\omega_nt)\right)$ is periodic in time. We point out that our analysis does not depend crucially on that especially $\gamma(t)$ is time-dependent. We could alternatively consider a constant $\gamma$ and a time-periodic Hamiltonian. 

We take (\ref{eq:Lindblad:general}) as our starting point, {\it i.e.} we take it as given. We may note, however, that some extra care is needed when deriving time-dependent LE's, especially for systems where the rate $\gamma(t)$ end up being time-dependent~\cite{dann2018time}. The LE, despite being explicitly time-dependent, preserves positivity and the trace of the density matrix, {\it i.e.} it has the aforementioned CPTP property~\cite{nielsen2002quantum}, something that will be important for the following discussions.

\subsection{Straightforward vectorisation of the master equation, and some properties of the Liouvillian} \label{sec:vectorisation}

Due to the linearity of the LE we can always rewrite it as a matrix equation. In this and the next subsection we present the most common ways how to parameterise the density operator and express it in terms of a vector. For larger dimensions, the method outlined in this section is typically favorable. In addition, it has advantages when it comes to numerical simulations since many numerical routines can be more or less straightforwardly applied. For smaller systems it is, however, often practical to consider a different parameterisation. 

By stacking either the columns or rows of the density matrix, we note that eq.~\eqref{eq:Lindblad:general} can be written as
\begin{equation} \label{eq:vectorized:Lindblad}
\frac{d}{d t}|\rho \rrangle  = \hat{\mathcal{L}}(t)  \kket{\rho}, 
\end{equation}
where  $\kket{\rho}$ is the vectorised version of the density matrix $\hat \rho$ and $\hat{\mathcal{L}}(t)$ is the {\it Liouvillian}  (sometimes also called {\it Lindbladian}\footnote{As Ingemar Bengtsson once formulated it: \textit{Here at the physics department in Stockholm we only have one researcher being a noun and that is G\"oran}}). The Louvillian is given by 
\begin{align} \label{eq:L:definition}
\hat{\mathcal{L}}(t) &= - i \bigl( \hat H(t) \otimes \mathds{1} -  \mathds{1}\otimes \hat H^{\rm{T}}(t) \bigr)  \\
& +  \frac{\gamma(t)}{2} \left[ 2 \hat L\otimes \hat L^{\dagger \rm{T}}\!-\! \hat L^\dagger \hat L \otimes \mathds{1}+ \mathds{1}\otimes  (\hat L^\dagger \hat L)^{\rm{T}}\! \right]. \nonumber 
\end{align}
The formal solution to the time-dependent LE becomes
\begin{equation} \label{eq:Lindblad:formal:solution}
\kket{\rho(t)} = \hat{\mathcal{S}}(t) \kket{\rho_0},
\end{equation}
where $\hat{\mathcal{S}}(t)$ is the time-ordered exponential of $\hat{\mathcal{L}}(t)$:
\begin{equation} \label{eq:formal:solution:S}
\hat{\mathcal{S}}(t) = \mathcal{T} \mathrm{exp} \left[ \int^t_0 \mathrm{d} t' \, \hat{\mathcal{L}}(t') \right]. 
\end{equation}
This is a key expression that captures both the unitary and the non-unitary evolution. 

For a pure state, $\hat \rho = \ketbra{\psi}$, the vectorised state reads
\begin{align}
\kket{\rho} = \ket{\psi} \otimes \ket{\psi^*}. 
\end{align}
The state $\kket{\rho}$ lives in the enlarged state space, {\it Liouville space}~\cite{gyamfi2020fundamentals}, comprising all pure and mixed states, but also {\it unphysical states}. Thus, a general vector $\kket{\nu}$ in this space does not have to represent a physical density matrix, meaning that it need not be positive semi-definite nor have a unit trace. For the Hilbert-Schmidt scalar product, in the Liouville space, $\bbraket{\hat\rho|\hat\varrho}=\mathrm{Tr}\left[\hat\rho\hat\varrho\right]$. Eigenvectors $\kket{\varrho_n}$ and eigenvalues $\mu_n$ of any time-independent Liouvillian $\hat{\mathcal{L}}$ are given by the standard expression
\begin{equation}\label{egekv}
    \hat{\mathcal{L}}\kket{\varrho_n}=\mu_n\kket{\varrho_n}.
\end{equation}
It follows that the steady state $\kket{\varrho_\mathrm{ss}}$ is the eigenvector with $\mu_n=0$. There can exist more than a single steady state, even though in most cases the steady state is unique for time-independent Liouvillians~\cite{spohn1977algebraic,hedvall2017dynamics}. Since $\hat{\mathcal{L}}$ is in general not a Hermitian operator (or rather $i\hat{\mathcal{L}}$ to be precise), we must distinguish between {\it left} and {\it right eigenvectors}; in (\ref{egekv}), for example, we consider the right eigenvectors~\cite{ashida2020non}. We may note that the left and right eigenvectors both form complete bases, hence any state can be expressed in terms of them, {\it e.g.}
\begin{equation}\label{exp}
    \kket{\rho}=\sum_nd_n\kket{\varrho_n}.
\end{equation}
These basis vectors are not, in general, orthogonal, {\it i.e.} $\bbraket{\varrho_n|\varrho_m}\neq\delta_{nm}$, while the left/right eigenvectors can indeed be made orthogonal, something referred to as {\it biorthogonality}~\cite{ashida2020non,brody2013biorthogonal,kunst2018biorthogonal}.

Given the expansion~(\ref{exp}), it follows that its time evolution reads
\begin{equation}
    \kket{\rho(t)}=\sum_nd_ne^{\mu_nt}\kket{\varrho_n}.
\end{equation}
Solving the dynamics by finding the eigenvalues of the Louvillian is also known as the {\it damping-basis approach}~\cite{briegel1993quantum}. 

Since the LE is trace-preserving we find the following important result for the eigenstates~\cite{minganti2019quantum}. Let $\hat\varrho_n$ be the density matrix corresponding to the eigenvector $\kket{\varrho_n}$, and $\hat\varrho_n(t)$ the time-evolved density matrix, it then follows
\begin{equation}\label{proof}
    \mathrm{Tr}\left[\hat\varrho_n\right]=\mathrm{Tr}\left[\hat\varrho_n(t)\right]=\mathrm{Tr}\left[\hat\varrho_n\right]e^{\mu_nt}.
\end{equation}
In the first line, we used the trace conservation, and in the second line, we used $\hat\varrho(t)=\hat\varrho \,e^{\mu_nt}$. Since $e^{\mu_nt}\neq1$, unless $\mu_n=0$, we must have that $\mathrm{Tr}\left[\hat\varrho_n\right]=0$. Thus, the only eigenvector representing a physical density matrix is the steady state.  {\textit All} other eigenvectors are traceless, and thereby unphysical~\cite{hedvall2017dynamics,minganti2019quantum}, which is the reason why we call them {\it eigenvectors} rather than {\it eigenstates} -- they do not represent states. This implies that for the expansion~(\ref{exp}) we can actually separate the sum into two parts
\begin{equation}
    \kket{\rho}=\kket{\varrho_\mathrm{ss}}+\sum_nd_ne^{\mu_nt}\kket{\varrho_n}.
\end{equation}
Moreover, to warrant positive semi-definiteness of the evolved state, the Liouvillian spectrum comes in complex pairs, {\it i.e.} for each complex $\mu_n$ there exists another eigenvalue $\mu_m=\mu_n^*$~\cite{minganti2018spectral,buvca2019non}.

The above discussion breaks down when the spectrum is degenerate. More precisely, upon varying some system parameter $\lambda$ and whenever any two or more eigenvalues coincide, $\mu_n(\lambda)=\mu_m(\lambda)$, the corresponding eigenvectors coalesce $\kket{\varrho_n}=\kket{\varrho_m}$, implying that the Liouvillian is not diagonalisable at this point~\cite{heiss2012physics,minganti2019quantum}. Such EPs mark a non-analytic behaviour in the spectrum typically characterized by some square-root singularity.

\subsection{Bloch vector vectroisation of the master equation}
In the quantum optics community, an alternative to the above vectorisation ~(\ref{eq:vectorized:Lindblad}) is often considered. This consists in parametrising the density matrix in terms of the {\it Bloch vector} ${\bf R}$  
\begin{equation}\label{dens}
\hat\rho=\frac{1}{D}\left(\mathbb{I}+\sqrt{\frac{D(D-1)}{2}}\mathbf{R}\cdot\lambda\right).
\end{equation}
Here, $D=\mathrm{dim}\{\mathcal{H}\}$ is the dmension of the Hilbert space, $\mathbf{R}=(r_1,\,r_2,\,\dots,\,r_{D^2-1})$ is the generalized Bloch vector, and the vector $\lambda=(\hat\lambda_1,\,\hat\lambda_2,\dots,\,\hat\lambda_{D^2-1})$ is composed of the generalized {\it Gell-Mann matrices}~$\hat\lambda_i$~\cite{hioe1981n}. The set of matrices $\hat\lambda_i$ matrices are Hermitian, traceless and together with the identity $\mathbb{I}$ they are mutually orthogonal, {\it e.g.} $\mathrm{Tr}\left[\hat\lambda_i\hat\lambda_j\right]=D\delta_{ij}$. Moreover, they form the generators of the group $SU(D)$. The Bloch vector length $|\mathbf{R}|\leq1$ such that the state space can be represented by a `Bloch hypersphere. However, whenever $D>2$ not all points within the Bloch sphere represent physical states~\cite{kimura2003bloch,goyal2016geometry}. 

The LE parametrised in terms of the Bloch vector takes the general form
\begin{equation}\label{meq}
\frac{\mathrm{d}}{\mathrm{d}t}\mathbf{R}=\mathcal{M}\mathbf{R}+\mathbf{b}.
\end{equation}
The advantage of this representation is that the constraint $\mathrm{Tr}[\hat\rho]=1$ removes one dimension of the matrix; the dimension of the {\it Bloch Liouvillian matrix} $\mathcal{M}$ is $(D^2-1)\times(D^2-1)$ while the dimension of $\hat{\mathcal{L}}$ is $D^2\times D^2$, which in higher dimension might not make a big difference. The price we pay is that eq. (\ref{dens}) is in general not homogeneous due to the {\it pump-term} ${\bf b}$. Note that when ${\bf b}=0$ a trivial steady state, ${\bf R}_\mathrm{ss}=0$ corresponds to the maximally mixed state. Typically, ${\bf b}=0$ when the Lindblad operators $\hat L_i$ are Hermitian~\cite{hedvall2017dynamics}. For a non-zero pump-term, we see that the steady-state Bloch vector becomes  ${\bf R}_\mathrm{ss}=-\mathcal{M}^{-1}{\bf b}$. 

As a final remark, we note that the pump term can be absorbed into $\mathcal{M}$ by considering a `Bloch vector' $\tilde{\bf R}=(\alpha,{\bf R})$, such that the density matrix is parametrised as $\hat\rho=\frac{1}{D}\left(\alpha\mathbb{I}+\sqrt{\frac{D(D-1)}{2}}\mathbf{R}\cdot\lambda\right)$, and the eq.~(\ref{meq}) becomes $\frac{\mathrm{d}}{\mathrm{d}t}\tilde{\mathbf{R}}=\tilde{\mathcal{M}}\tilde{\mathbf{R}}$. Hence, we have given up the normalisation constraint by introducing a new parameter $\alpha$, and the dimension of $\tilde{\mathcal{M}}$ is now $D^2\times D^2$. The voctorised density matrix $\kket{\rho}$ and the expanded Bloch vector $\tilde{\bf R}$ can be related by some matrix $V$, {\it i.e.} $\kket{\rho}=V|\tilde{\bf R}\rangle$, were we wrote the Bloch vector as a ket. Thus, the Liouvillian and extended Bloch Liouvillian are related via $\tilde{\mathcal{M}}=V^{-1}\hat{\mathcal{L}}V$. For the qubit
\begin{equation}
    V=\frac{1}{2}\left[
    \begin{array}{cccc}
    1 & 0 & 0 & 1\\
    0 & 1 & i & 0\\
    0 & -i & 1 & 0\\
    1 & 0 & 0 & -1
    \end{array}\right],    
\end{equation}
which up to a constant pre-factor is unitary, and hence the two spectra are equivalent (up to a factor). We may therefore talk about the ``Liouvillian spectrum'' without having to specify whether we use $\hat{\mathcal{L}}$ or $\mathcal{M}$. This does not, however, hold in higher dimensions.

\section{Analysis of a periodic two-level system} \label{sec:two:level:systems}

\subsection{Model system}
Considering a single qubit, we will stick to the Bloch representation~(\ref{dens}), rather than working with the vectorising outlined in subsection~\ref{sec:vectorisation}. For a qubit the Bloch vector takes the simple form $\mathbf{R}=(r_x,\,r_y,\,r_z)$, and $\lambda=(\hat\sigma_x,\,\hat\sigma_y,\,\hat\sigma_z)$ is composed by the Pauli matrices. In a proper frame, we may take a general two-level Hamiltonian as
\begin{equation} \label{eq:main:Hamiltonian}
\hat H_\mathrm{qu}=\left[
\begin{array}{cc}
\delta & g\\
g & -\delta
\end{array}\right],
\end{equation}
with $\delta$ and $g$ real. For multiple Lindblad operators $\hat L_i$, the Bloch equations reads
\begin{equation}
\partial_t\mathbf{R}=\mathcal{M}_H\mathbf{R}+\sum_i\left(\mathcal{M}_{L_i}\mathbf{R}+\mathbf{b}_i\right).
\end{equation}
where the anti-symmetric matrix
\begin{equation}
\mathcal{M}_H=2\left[
\begin{array}{ccc}
0 & -\delta & 0\\
\delta & 0 & -g\\
0 & g & 0
\end{array}\right], 
\end{equation}
derives from the unitary Hamiltonian evolution, and 
\begin{equation}
\mathcal{M}_{L_i}=\gamma_i(t)\left[
\begin{array}{ccc}
-\left(|L_{yi}|^2+|L_{zi}|^2\right) & L_{xi}L_{yi}^*+iL_{1i}L_{zi}^* & L_{zi}^*L_{xi}-iL_{1i}L_{yi}^*\\
L_{xi}^*L_{yi}-iL_{1i}L_{zi}^* & -\left(|L_{xi}|^2+|L_{zi}|^2\right) & L_{yi}L_{zi}^*+iL_{1i}L_{xi}^*\\
L_{zi}L_{xi}^*-iL_{1i}L_{yi}^* & L_{yi}^*L_{zi}-iL_{1i}L_{xi}^* & -\left(|L_{xi}|^2+|L_{yi}|^2\right)
\end{array}\right]+c.c.
\end{equation} 
is the contribution to $\mathcal{M}$ resulting from the jump operator $\hat L_i$, and the pump-term
\begin{equation}
\mathbf b_i=2i\gamma_i(t)\left[
\begin{array}{c}
L_{yi}^*L_{zi}\\
L_{zi}^*L_{xi}\\
L_{xi}^*L_{yi}
\end{array}\right]+c.c.\,.
\end{equation} 
The jump operators have been expressed in terms of the Pauli matrices; $\hat L_i=L_{1i}\mathbb{I}+L_{xi}\hat\sigma_x+L_{yi}\hat\sigma_y+L_{zi}\hat\sigma_z$. Here we see that Hermitian jump operators $\hat L_i$ generate vanishing pump-terms, {\it i.e.} $\mathbf b_i=0$. The pump-terms also disappear when only a single $L_{\alpha i}$ ($\alpha=x,\,y,\,z$) is non-zero, {\it e.g.} dephasing in a given basis. 

Let us consider an example where the pump term is zero, $\mathbf b_i$, for example, $\hat L=\hat\sigma_y$, which represents a dephasing in the $y$-basis. The Lindblad term can also be viewed as an incoherent pump, in addition to the coherent drive achieved for $g\neq0$. The Bloch Liouvillian matrix takes the form
\begin{equation}\label{m1}
    \mathcal{M}=\left[
    \begin{array}{ccc}
    -\gamma(t) & -\delta & 0\\
    \delta& 0 & -g\\
    0 & g & -\gamma(t)
    \end{array}\right].
\end{equation}
Using the Gell-Matrices $\hat\lambda_j$ ($i=1,\,2,\dots,\,8$), we may write $\mathcal{M}=-\gamma(t)\hat\lambda-i\delta\hat\lambda_2-i g\hat\lambda_7$, with $\hat\lambda=-2\mathbb{I}/3-\hat\lambda_3/2+\sqrt{3}\hat\lambda_8/6$. 

\subsection{Exceptional points and Bloch vector evolution}

It is worth analysing the adiabatic diagonalisation of~(\ref{m1}) to gain insight into the system's dynamics. Adiabaticity for open quantum systems differs in some respects from how one thinks about it for closed quantum systems~\cite{sarandy2005adiabatic,albash2015decoherence}. For example, the eigenvectors are typically not stationary states, and in particular for the Liouvillian $\hat{\mathcal{L}}$ any eigenvector with $\mathrm{Re}(\mu_n)<0$ will decay as time progresses. Hence, its norm is not preserved even for a time-independent Liouvillian. Moreover, turning to the Bloch Liouvillian matrix we should stress that the norm of the eigenvectors $\mathbf{R}_n$ ({\it i.e.} $\mathcal{M}\mathbf{R}_n=\nu_n\mathbf{R}_n$) plays a physical role; parallel Bloch vectors with different norms represent different density matrices $\hat\varrho_n$. Another aspect is the fact pointed out in Section~\ref{sec:vectorisation}, namely that eigenvectors of a NH matrix need not be orthogonal. Despite these complications, we can perform a time-instantaneous adiabatic diagonalisation of $\mathcal{M}(t)$. For the present example, we find the adiabatic eigenvalues
\begin{equation}\label{adeg}
    \begin{array}{l}
    \nu_0(t)=-\gamma(t),\\ \\
    \displaystyle{\nu_\pm(t)=-\frac{\gamma(t)}{2}\pm\sqrt{\frac{\gamma^2(t)}{4}-g^2-\delta^2}}.
    \end{array}
\end{equation}
We do not write out the explicit expressions for the Bloch eigenvectors ${\bf R}_{0,\pm}(t)$. 

\begin{figure}[t]
    \centering
    \includegraphics[width = 7.6cm]{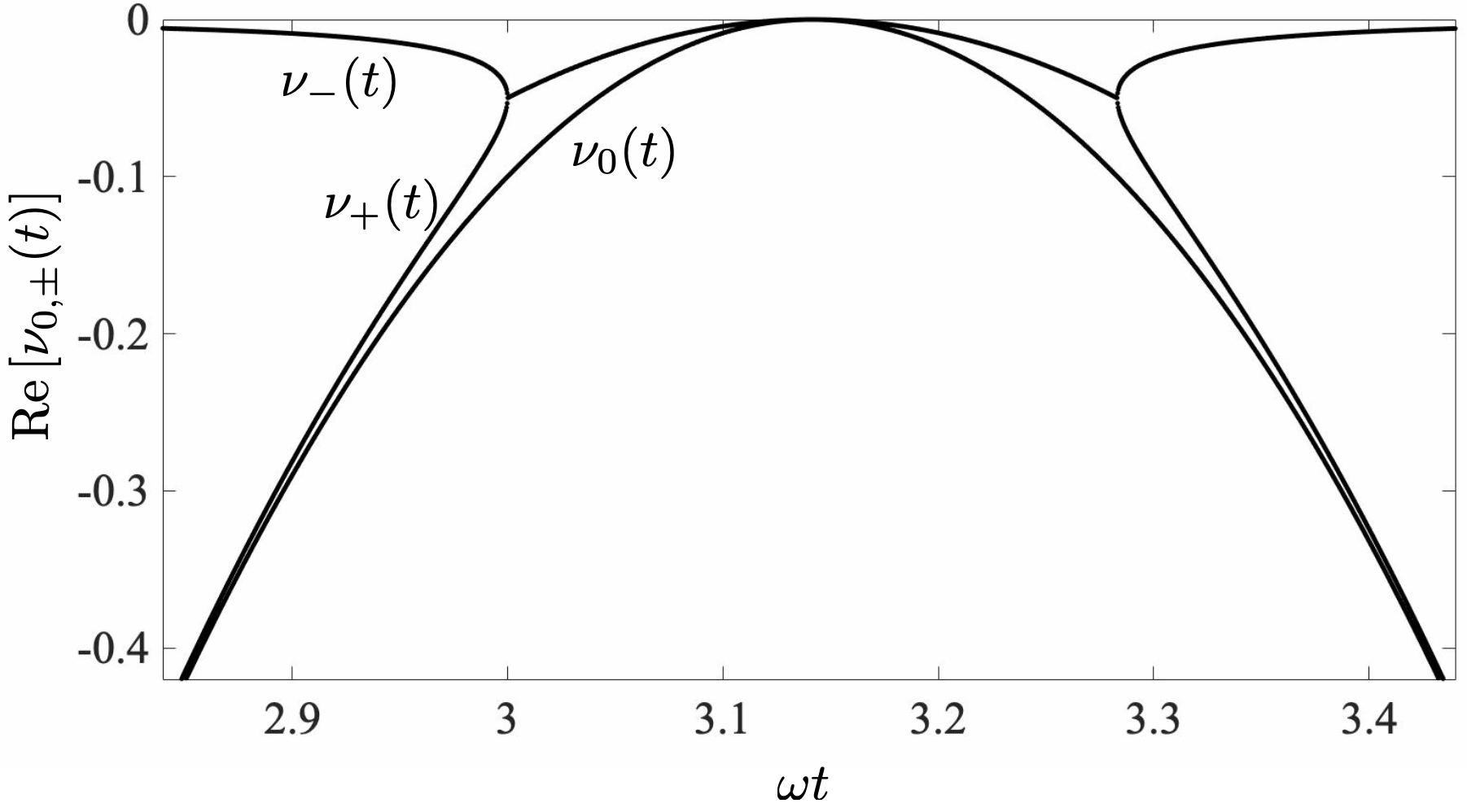}
    \caption[]{ The real parts of the adiabatic eigenvalues (\ref{adeg}) in the vicinity of a pair of EPs. The dimensionless parameters used are $\delta=0.05$, $g=0$, $\gamma_0=10$, and $\omega=0.05$. }
    \label{specfig}
\end{figure}

Using $\gamma(t)=\gamma_0\left[1+\cos(\omega t)\right]$ we have that for times $t_n$ such that $\omega t_n=\arccos\left(2\sqrt{\left(g^2+\delta2\right)}/\gamma_0-1\right)+2\pi n$, with $n\in\mathbb{N}$, the eigenvalues $\nu_\pm(t)$ coincide and EPs in terms of a square root singularity emerge. This is shown in fig.~\ref{specfig} where we zoom in around two EPs. Note that the EP results from `balancing' the term stemming from the Lindblad operator with those of the Hamiltonian~\cite{am2015exceptional}, and furthermore, the EP remains even if we let either $g$ or $\delta$ be zero.

The appearance of EPs in the Liouvillian may influence the evolution of the system. From the general discussion related to eq.~(\ref{proof}) we saw, however, that eigenvectors of the Liouvillian typically do not represent physical states, and this should, in one way or another, affect how EPs can be studied in real experiments. It is known that if we consider an EP of second order ({\it i.e.} doubly degenerate) and an initial instantaneous eigenvector, the vector will have swapped to the other eigenvector if the EP is adiabatically encircled~\cite{heiss2012physics,ashida2020non,minganti2019quantum}. This is a generalisation of the geometric phase to EPs; going around once swaps the vectors, going around twice adds an overall minus sign to the vector, going around three times causes the vectors to be swapped again but with a minus sign, and going around four rounds we return to the original vector. Recently, a qubit system exposed to both dephasing and dissipation was experimentally analysed within the Lindblad formalism, and especially the geometric properties upon encircling them were analysed~\cite{chen2022decoherence}.

Contrary to the situations above, let us focus on how the EPs affect the evolution of the system under a periodic modulation (for a time-dependent, but not periodic, study see ref.~\cite{khandelwal2021signatures}). In our case, this will cause the system to repeatedly experience the presence of EPs. Since the gap closes at an EP, and eigenvalues do not have continuous derivatives at these points, adiabaticity is expected to break down when the system is driven through any EP. However, on second thoughts, it is not as clear how adiabaticity is affected by the EPs. We expect that if we start in an initial eigenvector ${\bf R}_n(0)$ of $\mathcal{M}(0)$, for slow enough changes and large enough gaps, we expect that the system evolves along the instantaneous eigenstate ${\bf R}_n(t)$ with the prefactor $\exp\left(\int_0^t\,\nu_n(t')dt'\right)$ (we do not need to care about possible geometric phases for our argument~\cite{dasgupta2007decoherence}). Note that if we require the initial vector to represent a physical state, ${\bf R}_n(t)$ has to be a real vector and this implies that $\mathrm{Im}[\nu_n(0)]=0$. Thus, under adiabatic evolution the adiabatic vector ${\bf R}_n^{(\mathrm{ad})}(t)=\exp\left(\int_0^t\,\nu_n(t')dt'\right){\bf R}_n(t)$ stays real. This follows from the CPTP property of the LE. If the system is driven through an EP, after the EP the eigenvalue is no longer purely real and it comes with a companion complex eigenvalue (for the qubit case these are the $\nu_\pm(t)$ eigenvalues of eq.~(\ref{adeg})). Even so, the CPTP property implies that the evolved Bloch vector ${\bf R}(t)$ has to remain real, and hence, regardless of how slow or adiabatic the change is the vector has to become a superposition of the two companion instantaneous eigenvectors. Hence, there is no way the vector can remain in a single Bloch eigenvector if it traverses through EPs -- the CPTP property selects a linear combination that guarantees the positivity of the evolved state.

\begin{figure}[h!]
    \centering
    \includegraphics[width = 8.2cm]{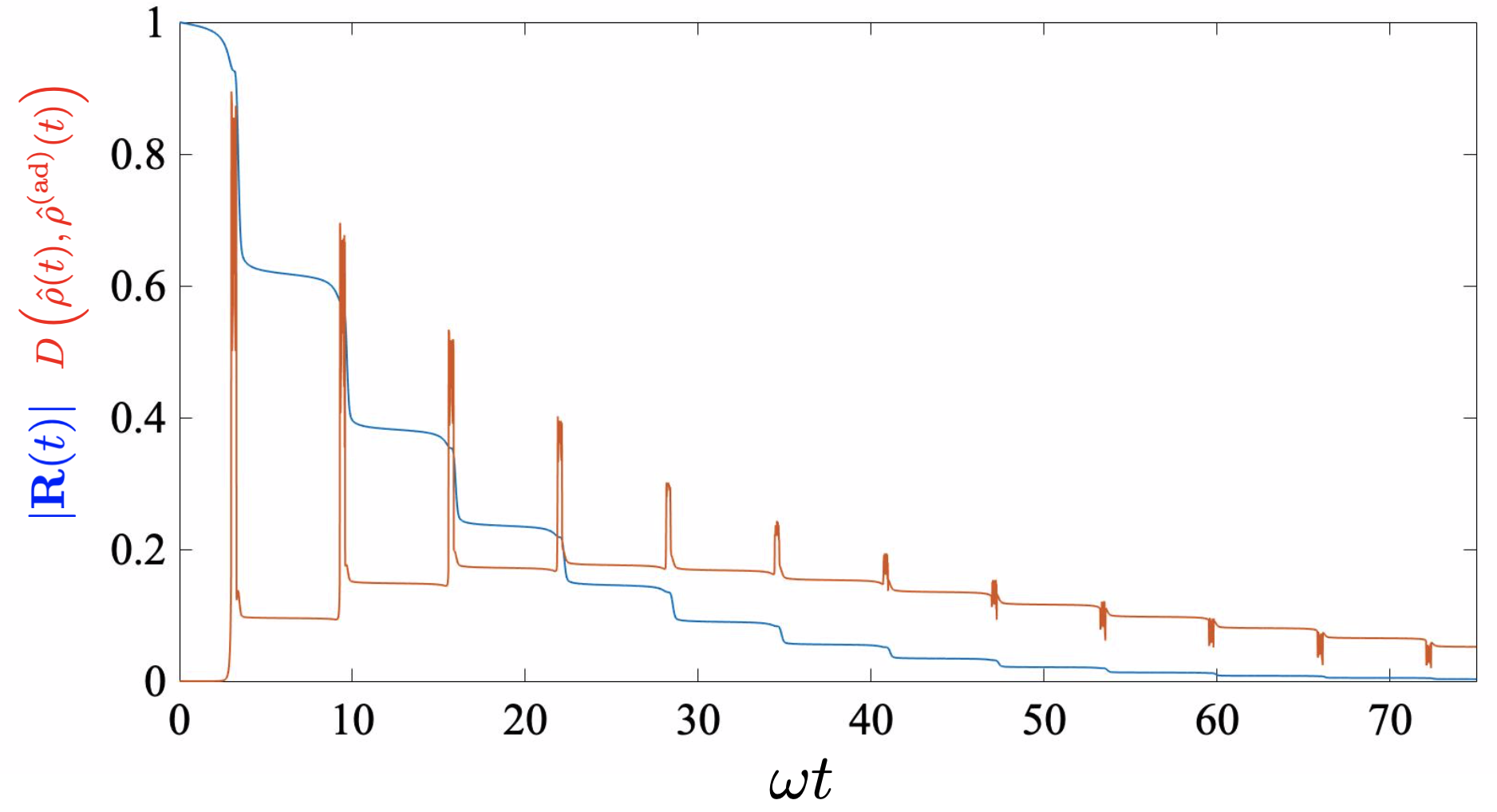}
    \caption[]{The evolution of the Bloch vector amplitude (blue solid line) and the trace distance~(\ref{trace}) (red solid line). For the given parameters, the adiabatic eigenvalues~(\ref{adeg}) displays pairs of EPs (see fig.~\ref{specfig}) where the Bloch vector length rapidly decreases and the trace distance drastically changes. In between the EPs, both the Bloch vector and the trace distance stay roughly constant, {\it i.e.} here the evolution is approximately adiabatic. At larger times, the Bloch vector approaches ${\bf R}\rightarrow{\bf R}_\mathrm{ss}=(0,0,0)$ which is the unique steady state for the system. The parameters are the same as those of fig.~\ref{specfig}, and the system is initialised in the eigenvector ${\bf R}_-(0)$. }
    \label{qubitevolution}
\end{figure}

The steady state of the Liouvillian matrix~(\ref{m1}) is the maximally mixed state $|{\bf R}_\mathrm{ss}|=0$, meaning that even for the periodic decay rate the system should eventually approach this state. The eigenvector ${\bf R}_0(t)$ has the largest negative real part of the eigenvalue and therefore shows the fastest instantaneous decay, while ${\bf R}_-(t)$ decays the slowest. Let us consider an initial state ${\bf R}(0)={\bf R}_-(0)$ and evolve it in time according to the Liouvillian~(\ref{m1}). To explore the influence of the EPs we pick parameters such that $\nu_\pm(t)$ coincide at the aforementioned points $t_n$. We wish to explicitly explore the adiabaticity and we thereby compare the evolved state with the adiabatic state ${\bf R}_-^{(\mathrm{ad)}}(t)$. In particular, we calculate the trace distance between the corresponding density matrices
\begin{equation}\label{trace}
    T(\hat\rho,\hat\rho^{(\mathrm{ad})})=\frac{1}{2}\mathrm{Tr}\left|\hat\rho(t)-\hat\rho^{(\mathrm{ad})}(t)\right|.
\end{equation}
Over longer times, the trace distance will approach zero since the two Bloch vectors vanish in the steady state. Nevertheless, it gives a rough idea of how well the evolved state follows the adiabatic one.

In fig.~\ref{qubitevolution} we display both the trace distance and the purity of the state given by the Bloch vector amplitude $|{\bf R}(t)|$. For the given parameters, same as for fig.~\ref{specfig}, one finds two EPs nearby in time, and in these instances, the purity suddenly drops. The purity is monotonously decreasing so we see no signs of reestablished coherences, which could possibly happen for a time-dependent LE. The trace distance demonstrates that the breakdown of adiabaticity occurs at the EPs for these parameters; apart from the vicinities of the EPs the trace distance stays approximately constant. For other parameters, where no EPs occur, we see a qualitatively different behaviour with an approximate exponentially decaying purity. Thus, the EPs may influence the evolution of the system by making the dephasing highly non-exponential. In the next subsection, we will see that the emergence of EPs seems to have a more hidden effect which becomes clear by analysing the model in a Floquet basis.

We have observed that the appearance of EPs in the Bloch Liouvillian does not occur for just any Hamiltonians and Lindblad jump operators. Especially, in many solvable models one notices that the underlying Lie algebra is rather simple ({\it i.e} it is closed and has a low dimension), and it seems that these models have in common that they lack EPs. As an example, in ref.~\cite{rau2002embedding}, another set of Bloch equations than those of eq.~(\ref{m1}) were studied, and analytical solutions could indeed be given thanks to a closed Lie algebra. For our model, the set $\left\{\hat\lambda,\hat\lambda_3,\hat\lambda_8\right\}$ does not alone form a closed algebra. To try and further understand the underlying structure giving rise to EPs in two-level systems, we turn to a Lie algebra method first developed by Wei and Norman \cite{wei1963lie}. This method has been frequently used in the literature to solve the dynamics of both two-level \cite{scopa2019exact} and continuous variable systems \cite{qvarfort2019enhanced, qvarfort2020time}. The premise of the method is to identify a finite-dimensional Lie algebra that generates the time evolution of the system. Once such an algebra has been identified, a set of differential equations can be derived which fully determine the dynamics. 

We here briefly outline the method. Consider a Hamiltonian
\begin{equation} \label{eq:decoupling:Hamiltonian}
\hat H(t) = \sum_j^m G_j(t) \hat H_j  ,
\end{equation}
where $G_j(t)$ are (possibly) time-dependent coefficient, and the $\hat H_j$'s encode either free evolution or interaction terms. We then identify the Lie algebra $L$ by commuting the Hamiltonian operator terms $\hat H_j$. Whenever they produce a new operator that was not part of the initial set $\{ \hat H_j\}$, {\it e.g.} $[\hat H_i, \hat H_j] \propto \hat H_k$, we add it to the set. If a finite number of $n \geq m $ operators can be identified in this way, we have obtained a finite-dimensional Lie algebra that generates the dynamics. We define the time-evolution operator in the usual way as
\begin{align} \label{eq:U:time:ordered}
    \hat U(t) = \mathcal{T} \mathrm{exp}\left[ - \frac{i}{\hbar} \int^t_0 \mathrm{d}t^\prime \, \hat H(t^\prime) \right], 
\end{align}
where $\mathcal{T}$ indicates time-ordering. It is often difficult to derive a closed-form solution for $\hat U(t)$, but according to \cite{wei1963lie}, we state an ansatz with the help  of the factorised unitary
\begin{align} \label{eq:ansatz}
\hat U(t) &= \exp[ - i \, F_1(t) \, \hat H_1] \, \exp[ - i \, F_2(t) \, \hat H_2] \cdots   \exp[ - i \, F_n (t) \, \hat H_n ] ,
\end{align}
where $F_j $ are time-dependent coefficients and $\hat H_j$ are the elements of the Lie algebra. By then differentiating both~\eqref{eq:ansatz} and~\eqref{eq:U:time:ordered}, then equating the results, we can derive a set of differential equations 
\begin{align} \label{eq:matrix:form:equation}
\vec{G} = - i \,  \boldsymbol{\xi}  \, \dot{\vec{F}} , 
\end{align}
where $\vec{G}$ is the vector of Hamiltonian coefficients and $\vec{F}$ is the vector of corresponding coefficients defined in the ansatz~\eqref{eq:ansatz}. 

The Lie algebra decoupling method may be extended to open system dynamics by treating the vectorised Louvillinan~\eqref{eq:vectorized:Lindblad} as the Hamiltonian and stating an ansatz for the time-evolution in the form of 
\begin{align}
    \hat{\mathcal{S}}(t) &= \mathrm{exp}[ D_1 \hat{\mathcal{L}}_1] \, \mathrm{exp}[ D_2 \hat{\mathcal{L}}_2] \ldots  \mathrm{exp}[D_n \hat{\mathcal{L}}_n], 
\end{align}
where $D_j$ are time-dependent coefficients. 

Through this method, we may study the underlying structure of the algebra that generates the time evolution. 
After vectorising the Lindblad equation for the Hamiltonian~\eqref{eq:main:Hamiltonian} as well as the Lindblad operators $\hat L = \hat \sigma_y$, we obtain the terms~\footnote{Technically, the additional transpose that arises from our choice of vectorisation gives us the term $\hat \sigma_y \otimes \hat \sigma_y^*$, however, we note that $\hat \sigma_y^* = -\hat \sigma_y$. Thus we can consider the full Lie algebra without the transpose. }
\begin{align}
    &\hat \sigma_z \otimes \mathds{1} && \mathds{1} \otimes \hat \sigma_z && \hat \sigma_y \otimes \hat \sigma_y\nonumber \\
    &\hat \sigma_y \otimes \mathds{1} && \mathds{1} \otimes \hat \sigma_y 
\end{align}
By commuting the terms, we obtain the full Lie algebra that generates the evolution
\begin{align} \label{eq:full:algebra}
&\hat \sigma_x \otimes \mathds{1} &&\mathds{1} \otimes \hat\sigma_x && \hat\sigma_z \otimes \hat    \sigma_z \nonumber \\
& \hat \sigma_y \otimes \mathds{1} && \mathds{1} \otimes \hat\sigma_y && \hat\sigma_y \otimes \hat\sigma_y \nonumber \\
&\hat \sigma_z \otimes \mathds{1} && \mathds{1} \otimes \hat\sigma_z  && \hat\sigma_x \otimes \hat\sigma_x \nonumber  \\
&\hat\sigma_x \otimes \hat\sigma_y && \hat\sigma_y \otimes \hat\sigma_x && \hat\sigma_z \otimes \hat\sigma_y \nonumber \\
&\hat\sigma_x \otimes \hat\sigma_z  && \hat\sigma_z \otimes \hat\sigma_x && \hat\sigma_y \otimes\hat \sigma_z     
\end{align} 
This algebra contains all $15$ elements, and thus the ansatz for the non-unitary dynamics contains an equal number of terms. We conclude that despite the rather simple model used in our numerics $\hat HH\propto\hat\sigma_z$ and $\hat L=\hat\sigma_y$, it generates a high dimensional Lie algebra. 

Instead, we may consider a Hamiltonian and Lindblad operator that together generate a smaller algebra. In \cite{scopa2019exact}, it was shown that the following choice of Hamiltonian and Lindblad operators generates an algebra with only three elements:
\begin{align}
    \hat{\mathcal{H}} = - \frac{1}{2}\Omega \hat \sigma_z ,  \nonumber 
\end{align}
and Lindblad operators
\begin{align}
    &\hat L_1 = \hat \sigma_+,  && \hat L_2 = \hat \sigma_-,   && \hat L_3 = \hat \sigma_z . 
\end{align}
The ansatz reads
\begin{align} \label{eq:small:algebra:ansatz}
    \hat{\mathcal{S}}(t) = e^{F_z(t) \hat{\mathcal{H}}_z} \, e^{ F_\uparrow \hat{\mathcal{D}}_\uparrow }\, e^{F_\downarrow \hat{\mathcal{D}}_\downarrow} \, e^{F_{33} \hat{\mathcal{D}}_{33}}, 
\end{align}
where we have introduced 
\begin{align}
    \hat{\mathcal{H}}_3 =  - i \left( \mathds{1}_2 \otimes \hat \sigma_z - \hat \sigma_z^* \otimes \mathds{1}_2 \right), 
\end{align}
as well as the super-operators
\begin{align}
    \hat{\mathcal{D}}_{\uparrow ; \downarrow} = \frac{1}{2} \left( \hat{\mathcal{D}}_{11} + \hat{\mathcal{D}}_{22} \mp  i \hat{\mathcal{D}}_{12} \pm \hat{\mathcal{D}}_{21}\right), 
\end{align}
for which 
\begin{align}
    [\hat{\mathcal{D}}_\uparrow, \hat{\mathcal{D}}_\downarrow] = \hat{\mathcal{D}}_\uparrow - \hat{\mathcal{D}}_\uparrow,   
\end{align}
and where 
\begin{align}
    \mathcal{D}_{jk}= \hat \sigma_k^* \otimes \hat \sigma_j - \frac{1}{2}\mathds{1}_2 \otimes \hat \sigma_k \hat \sigma_j - \frac{1}{2} \hat \sigma_j^* \hat \sigma_k^* \otimes \mathds{1}_2. 
\end{align}
The differential equations for the coefficients in~\eqref{eq:small:algebra:ansatz} are given in \cite{scopa2019exact}. 

The link between the Hilbert space time-evolution $\hat{\mathcal{S}}(t)$ and the eigenvalues of the spectrum studied is that the eigenvalues of the Louvillian are given by 
\begin{align}
    \mathrm{diag} \hat{\mathcal{S}}(t) = \mathrm{diag}( e^{\lambda_1}, e^{\lambda_2}, e^{\lambda_3}, e^{\lambda_4}).
\end{align}
However, when analysing the spectrum of the Louvillian that yields the smaller algebra, we indeed find that no exceptional points arise. The lack of exceptional points for a smaller algebra indicates that we require a Hamiltonian and noise operators that generate the full SU(4) algebra to observe this phenomenon. At this point, this is just an observation and we have no proof for it.


\subsection{Floquet analysis -- Wannier-Stark ladders and parameter sensitivity}
To make the analog to Floquet theory~\cite{shirley1965solution,sambe1973steady,eckardt2017colloquium} more transparent, let us express the Bloch equations in terms of a Schr\"odinger-like equation with a NH Hamiltonian
\begin{equation}
    i\frac{\partial}{\partial t}|{\bf R}(t)\rangle=\mathcal{H}(t)|{\bf R}(t)\rangle,
\end{equation}
where $\mathcal{H}(t)=i\mathcal{M}(t)$ and $|{\bf R}(t)\rangle$ is the Bloch vector written as a ket. The periodicity of $\mathcal{H}(t)$ allows us to introduce a time-independent {\it Floquet Hamiltonian} $\mathcal{H}_F$. A method to find the corresponding Floquet Hamiltonian is to make use of the infinite {\it Euclidean algebra}~\cite{klimov2009group} 
\begin{equation}\label{nhs}
    \left[\hat E_0,\hat E\right]=-\hat E,\hspace{1cm}\left[\hat E_0,\hat E^\dagger\right]=\hat E^\dagger,\hspace{1cm}\left[\hat E,\hat E^\dagger\right]=0.
\end{equation}
We have the eigenstates of $\hat E_0$, $\hat E_0|m\rangle=m|m\rangle$, for $m\in\mathbb{Z}$, and ladder operators $\hat E|m\rangle=|m-1\rangle$ and $\hat E^\dagger|m\rangle=|m+1\rangle$. We further introduce the {\it phase states}, expressed in the {\it Floquet states} $|m\rangle$ as 
\begin{equation}
    |\theta\rangle=\lim_{N\rightarrow\infty}\sum_{m=-N}^Ne^{-i\theta m}|m\rangle,
\end{equation}
such that $\hat E|\theta\rangle=e^{i\theta}|\theta\rangle$ and $\hat E^\dagger|\theta\rangle=e^{-i\theta}|\theta\rangle$. 

Consider now a rotating frame with respect to $\omega\hat E_0$, {\it i.e.} $|\theta(t)\rangle=e^{-i\hat \omega E_0t}|\theta\rangle$. 
A practical feature is that $\langle\theta(t)|\hat E|\theta(t)\rangle=e^{-i(\omega t+\theta)}$, which implies that we can express the Hamiltonian in~(\ref{nhs}) as the expectation value of the time-independent Floquet Hamiltonian; $\mathcal{H}(t)=\langle\theta(t)|\mathcal{H}_F|\theta(t)\rangle$, and where we for simplicity have taken the parameter $\theta=0$. Thus, we have transformed the time-dependent problem of $\mathcal{H}(t)$ into a time-independent one of $\mathcal{H}_F$. In general, we can use the mapping
\begin{equation}
    \gamma(t)=\gamma_0\left[1+\cos(\omega t)\right]\rightarrow\frac{\gamma_0}{2}\left(2+\hat E+\hat E^\dagger\right)
\end{equation}
to go from $\mathcal{H}$ to $\mathcal{H}_F$. The Floquet Hamiltonian then becomes
\begin{equation}\label{fham}
\mathcal{H}_F=\left[\omega\hat E_0-i\frac{\gamma_0}{2}\left(2+\hat E+\hat E^\dagger\right)\hat\lambda-i\delta\hat\lambda_2-ig\hat\lambda_7\right],
    \end{equation}
where the $\lambda_\alpha$-operators are the Gell-Mann matrices. 

\begin{figure}[h!]
    \centering
    \includegraphics[width = 8.2cm]{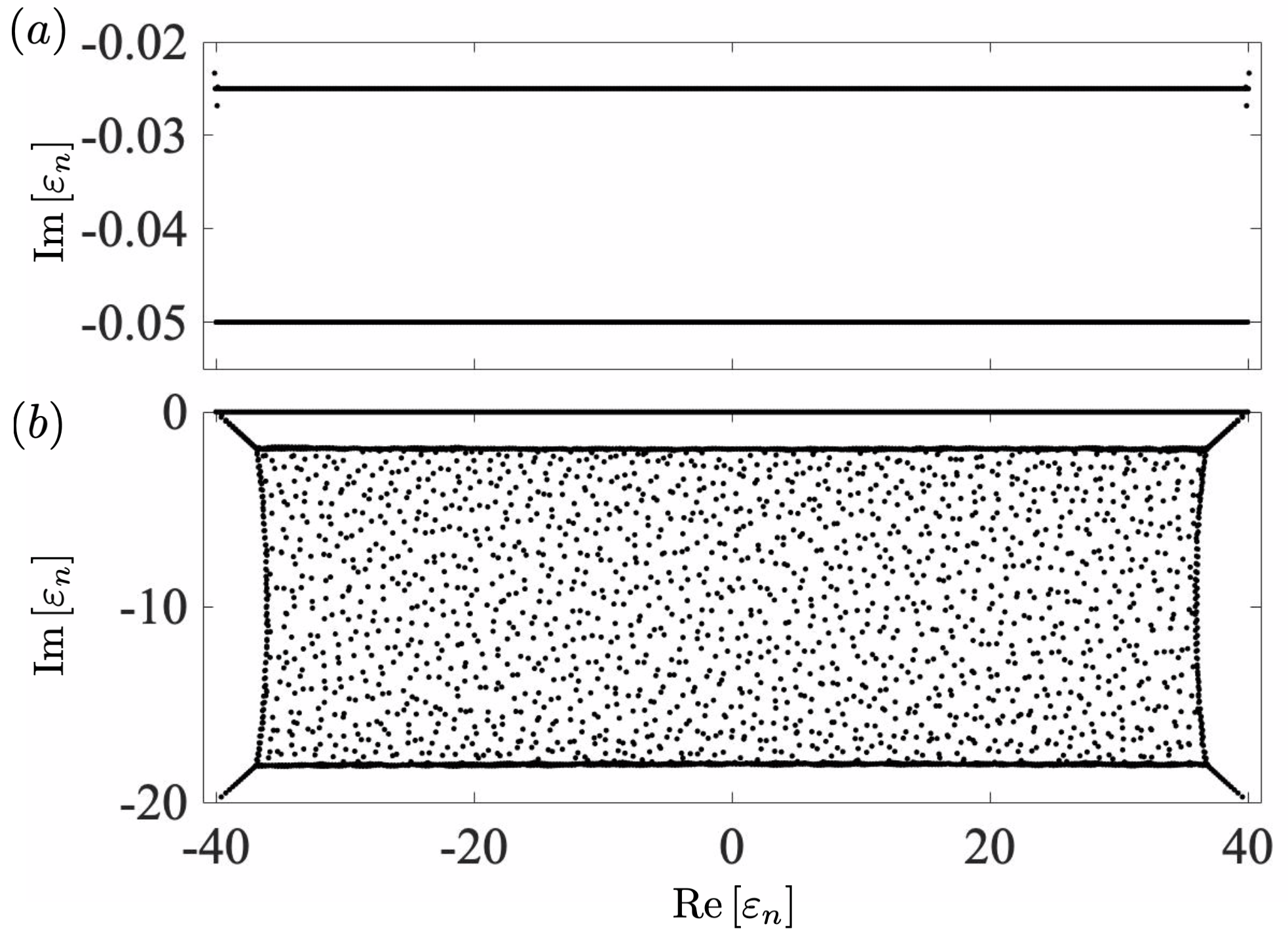}
    \caption[]{The numerically extracted Floquet spectrum $\varepsilon_n$ for two different set of parameters; ($a$) $\delta=0.05$, $g=0$, $\gamma_0=0.1$, and $\omega=0.05$ and ($b$) $\delta=0.05$, $g=0$, $\gamma_0=05$, and $\omega=0.05$ (same as in Figs.~\ref{specfig} and~\ref{qubitevolution}). We truncate the Hamiltonian to $-800\leq m\leq800$. The upper plot demonstrates the three Wannier-Stark ladders (two of them being degenerate). At the edges of the spectrum, there are a few scattered eigenvalues, but it should be understood that these are artifacts from having a truncated Hamiltonian; in the thermodynamic limit, the real parts of the Wannier-Stark ladders extend from $-\infty$ to $+\infty$. In the lower plot, we consider parameters well in the regime where EPs appear in the Liouvillian (see fig.~\ref{specfig}). Here something interesting happens; a fraction of the energies depart from the Wannier-Stark ladders and start to fill the space between the two lower ladders.}
    \label{floquetspec}
\end{figure}

Apart from being NH, due to the presence of the imaginary $i$, and unbounded from below, the Hamiltonian~(\ref{fham}) bears many similarities with other known models. The ladder operators $\hat E^\dagger$ and $\hat E$ can be seen as re-scaled bosonic creation/annihilation operators, and then the Hamiltonian is akin an $SU(3)$ {\it Jaynes-Cummings} or quantum {\it Rabi model}~\cite{larson2021jaynes}. Here, the {\it bare states} takes the form $|m,\hat e_k\rangle$ (with $\hat e_k$ representing the unit vector for the three Bloch vectors with components $k=x,\,y,\,z$).  Alternatively, the Hamiltonian also describes a single-particle tight-binding lattice model. More precisely we find a 1D three-legged ladder lattice, with the Floquet quantum number $m$ denoting the $m$'th rung sites in the lattice, the ladder operators describe (complex) hopping along the legs, while the $\hat\lambda_\alpha$-operators induce hopping along the rungs instead. Similar lattice models arising in quantum optical models were recently studied in~\cite{saugmann2022fock}. The {\it bare} term $\omega\hat E_0$ plays the role of a tilt of the lattice or a constant force parallel to the lattice. A tilted tight-binding lattice model is known as a {\it Wannier-Stark ladder}~\cite{wannier1960wave,hartmann2004dynamics}. The characteristic properties of a (Hermitian) Wannier-Stark ladder are that the spectrum, in the thermodynamic limit, $E_m=\omega m$ is equidistant, and the $m$'th (Wannier-Stark) eigenstate is localized to the $m$'th site -- so-called {\it Stark localization}~\cite{emin1987existence,schulz2019stark}. Using knowledge from Jaynes-Cummings-like and Wannier-Stark models we can now analyse the properties of our Floquet Hamiltonian.

Figure~\ref{floquetspec} presents two examples of the numerically extracted spectrum $\varepsilon_n$ of the Floquet Hamiltonian $\mathcal{H}_F$,
\begin{equation}
    \mathcal{H}_F|\varphi_n\rangle=\varepsilon_n|\varphi_n\rangle,
\end{equation}
in the complex plane, one where the Liouvillian $\mathcal{M}(t)$ does not support EPs ($a$), and one were it does ($b$). The Floquet Hamiltonian was diagonalised in the bare basis $|m,\hat e_k\rangle$ with a truncation $-800\leq m\leq800$. We find that for small $\gamma_0$, the spectrum becomes what one could expect for a (complex) Wannier-Stark ladder, namely
\begin{equation}
    \varepsilon_{m,\{0,\pm\}}=\omega m+i\nu_{\{0,\pm\}},
\end{equation}
with $\nu_{\{0,\pm\}}=nu_{\{0,\pm\}}(t=0)$, and as visualised in ($a$). In this example, we find three Wannier-Startk ladders, of which two are degenerate. As soon as $\gamma_0$ is increased beyond the EP, the two degenerate ladders split. 

A novel phenomenon emerges when we increase $\gamma_0$ further, see ($b$). After a certain dephasing rate $\gamma_0$ ($\approx1$ for these parameters), two of the ladders start to break up; some eigenvalues pull out from the ladders and start to fill up the space between the two ladders. Recall, that in reality $-\infty<m<\infty$ meaning that the behaviour at the edges of the spectrum ({\it i.e.} large values of $|\mathrm{Re}\left[\varepsilon_n\right]|$ in the plot) is a finite size effect from the numerical truncation. This effect of collapsing the ladders gets stronger the larger $\gamma_0$ considered. Using an argument of analytic continuation to the complex plane of a regular Wannier-Stark ladder, it is tempting to think that this is a numerical artifact from diagonalising large NH matrices~\cite{budich2020non,edvardsson2022sensitivity}. We find no evidence of such numerical sensitivity by varying system sizes and numerical precision, but, on the other hand, we lack a mathematical proof that this indeed is a physical result. In recent years, {\it exponential sensitivity} of the spectrum, {\it i.e.} small perturbations $\epsilon$ render a spectral change scaling as $\sim\exp(\epsilon L)$ where $L$ is the spatial size of the perturbation ({\it e.g.} if the perturbation changes the boundary conditions $L$ is the system size), has been thoroughly studied in the realm of NH quantum mechanics where it is known to be related to the skin effect. The NH skin effect is a phenomenon in which the (left/right) eigenvectors of a lattice model become exponentially localized to one edge of the system~\cite{budich2020non,edvardsson2022sensitivity,okuma2020topological,zhang2022review}. 

\begin{figure}[h!]
    \centering
    \includegraphics[width = 8.2cm]{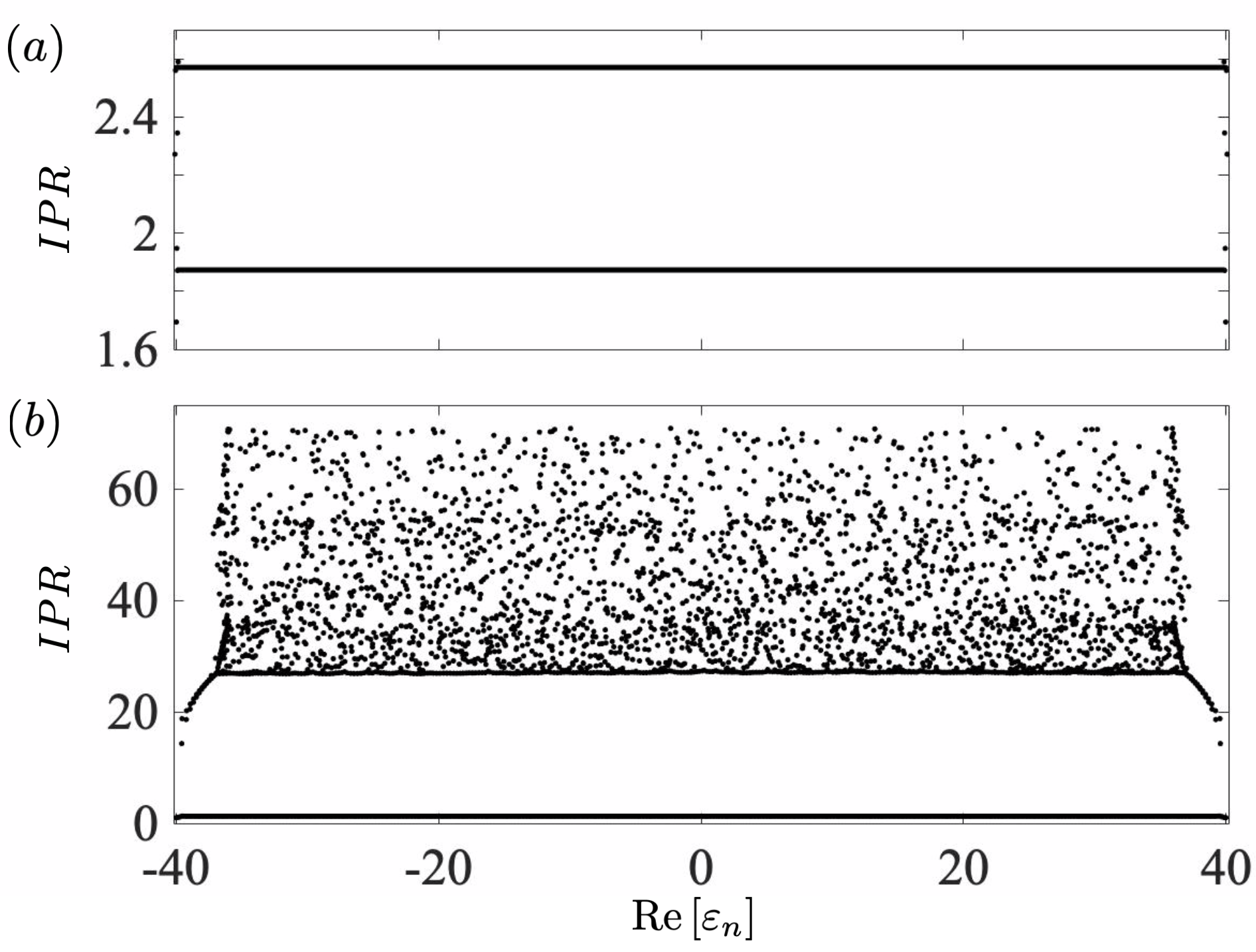}
    \caption[]{The $IPR$~(\ref{ipr}) vs. the real parts of the spectrum for the examples of fig.~\ref{floquetspec}. In the case of well-resolved Wannier-Stark ladders, we also find nicely localized eigenstates. In the other case when the ladders start to break up, we find much less localized states, especially for those states not within a ladder.}
    \label{iprfig}
\end{figure}

In our NH Wannier-Stark model, we indeed find that the scattered points display exponential sensitivity, both to parameter changes and to system size. There is an important aspect to point out that makes our model different from other NH models displaying exponential sensitivity, namely that our eigenstates are spatially localized. The other paradigm example in physics of extended (non-interacting) systems with localized eigenstates is disordered ones, which give rise to {\it Anderson localization}~\cite{kramer1993localization}. In disordered one-dimensional systems, all eigenstates become exponentially localized. One can imagine that Anderson localization and the skin effect localization could potentially conflict. Typically, in a system showing the skin effect any initial state will propagate towards the boundary of the system, but if conductivity vanishes due to Anderson localization it is unclear which effect will win. This was analysed in~\cite{jiang2019interplay}, and it was found that there is a trade-off between the disorder strength and the drift in the system determining which mechanism survives. In our model, the states are localized due to the Stark tilt, and there are no boundaries in the system. As a result, we may expect that the Stark localization should prevail, which is also what we find. To study the sensitivity we introduce the {\it inverse partition ratio}
\begin{equation}\label{ipr}
    IPR_n=\frac{1}{\sum_{m,k}\left|c_{m,k}^{(n)}\right|^4},
\end{equation}
where the $c_{m,k}^{(n)}$ are the coefficients of the $n$'th (right) eigenstate expanded in the bare states $|m,\hat e_k\rangle$, {\it i.e.} $|\varphi_n\rangle=\sum_{m,k}c_{m,k}^{(n)}|m,\hat e_k\rangle$. For a delocalized state, the $IPR_n$ scales with the system dimension, while for a fully localized state (occupying a single site) $IPR_n=1$. The $IPR$ for the examples of fig.~\ref{floquetspec} are shown in fig.~\ref{iprfig}. We present the $IPR$ as a function of the real part of the eigenvalue. When the system displays no exponential sensitivity we also find localized Stark-states (a). When the sensitivity has set in, the states of the Stark ladders get less localized. More precisely, numerics suggest that these states stay localized but instead of being exponentially localized, they become more Gaussian-like. We have explored whether there is a continuous phase transition governing this new behaviour, but we find no evidence for this. It is further unclear if this should be called a {\it Floquet skin effect}; the states stay localized, only a fraction of the states show the sensitivity, and there is no real boundary for the states to localize around.


\section{Conclusions} \label{sec:conclusion}

In this study, we focused on topics inspired by G\"oran Lindblad's interests, including open quantum systems, Lie algebras, and quantum dynamics. While exponential sensitivity and exceptional points (EPs) have been widely explored in non-Hermitian (NH) quantum mechanics, their investigation in the context of Lindblad equations (LEs) has been relatively scarce. To shed light on this issue, we investigated a periodically driven dephased qubit system that is simple yet non-trivial. Our results indicate that in the presence of strong dephasing, the dephasing evolution is no longer exponential. Instead, the system exhibits a step-like decay of the qubit's purity due to the presence of EPs. Particularly, in the vicinity of the EPs, adiabaticity and exponential decay break down, leading to sudden jumps of the Bloch vector.

We also explored the presence of EPs by examining the underlying Lie algebra of the system. We found that all 15 Lie algebra elements are necessary to close the algebra. Although other choices of Lindblad jump operators could be used, resulting in a lower-dimensional algebra~\cite{rau2002embedding,scopa2019exact}, they do not support EPs. It appears as if the dimension of the algebra cannot be too low for EPs to appear.

Finally, we derived a non-Hermitian Floquet Hamiltonian using the periodicity of the Liouvillian and analyzed its spectrum. As expected, we found a complex Wannier-Stark spectrum with localized eigenvectors. However, we also discovered new solutions that arise when the dephasing becomes stronger. These solutions are still localized, but they do not form a ladder spectrum and are exponentially sensitive to parameter changes. Similar sensitivity has been reported in the context of the skin effect. While we have explored the possibility of this effect being a numerical problem, we could not find any evidence for it. Nonetheless, a deeper understanding of this effect would be beneficial, but we leave it for future investigations.

\section*{Acknowledgements} 

JL posthumously acknowledges G\"oran Lindblad for many interesting discussions, especially during a reading course on {\it Paradoxes in Quantum Mechanics}. S.Q.~was funded in part by the Wallenberg Initiative on Networks and Quantum Information (WINQ) and in part by the Marie Skłodowska--Curie Action IF programme Nonlinear optomechanics for verification, utility, and sensing -- Grant-No.~101027183. Nordita is partially supported by NordForsk.

\bibliographystyle{unsrt}
\bibliography{td-therm}



\end{document}